\documentclass[twoside,twocolumn,9pt]{article}
\usepackage{extsizes}
\usepackage[super,sort&compress,comma]{natbib} 
\usepackage[version=3]{mhchem}
\usepackage[left=1.5cm, right=1.5cm, top=1.785cm, bottom=2.0cm]{geometry}
\usepackage{balance}
\usepackage{mathptmx}
\usepackage{sectsty}
\usepackage{graphicx} 
\usepackage{lastpage}
\usepackage[format=plain,justification=justified,singlelinecheck=false,font={stretch=1.125,small,sf},labelfont=bf,labelsep=space]{caption}
\usepackage{float}
\usepackage{fancyhdr}
\usepackage{fnpos}
\usepackage[english]{babel}
\addto{\captionsenglish}{%
  
}
\usepackage{array}
\usepackage{droidsans}
\usepackage{charter}
\usepackage[T1]{fontenc}
\usepackage[usenames,dvipsnames]{xcolor}
\usepackage{setspace}
\usepackage[compact]{titlesec}
\usepackage{hyperref}
\usepackage{bm}
\usepackage{amsmath,amssymb}

\usepackage{epstopdf}

\definecolor{cream}{RGB}{222,217,201}

\begin{document}

\pagestyle{fancy}
\thispagestyle{plain}
\fancypagestyle{plain}{
\renewcommand{\headrulewidth}{0pt}
}

\makeFNbottom
\makeatletter
\renewcommand\LARGE{\@setfontsize\LARGE{15pt}{17}}
\renewcommand\Large{\@setfontsize\Large{12pt}{14}}
\renewcommand\large{\@setfontsize\large{10pt}{12}}
\renewcommand\footnotesize{\@setfontsize\footnotesize{7pt}{10}}
\makeatother

\renewcommand{\thefootnote}{\fnsymbol{footnote}}
\renewcommand\footnoterule{\vspace*{1pt}%
\color{cream}\hrule width 3.5in height 0.4pt \color{black}\vspace*{5pt}} 
\setcounter{secnumdepth}{5}

\makeatletter 
\renewcommand\@biblabel[1]{#1}            
\renewcommand\@makefntext[1]%
{\noindent\makebox[0pt][r]{\@thefnmark\,}#1}
\makeatother 
\renewcommand{\figurename}{\small{Fig.}~}
\sectionfont{\sffamily\Large}
\subsectionfont{\normalsize}
\subsubsectionfont{\bf}
\setstretch{1.125} 
\setlength{\skip\footins}{0.8cm}
\setlength{\footnotesep}{0.25cm}
\setlength{\jot}{10pt}
\titlespacing*{\section}{0pt}{4pt}{4pt}
\titlespacing*{\subsection}{0pt}{15pt}{1pt}

\fancyfoot{}
\fancyfoot[LO,RE]{\vspace{-7.1pt}\includegraphics[height=9pt]{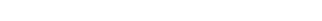}}
\fancyfoot[CO]{\vspace{-7.1pt}\hspace{13.2cm}\includegraphics{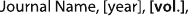}}
\fancyfoot[CE]{\vspace{-7.2pt}\hspace{-14.2cm}\includegraphics{head_foot/RF}}
\fancyfoot[RO]{\footnotesize{\sffamily{1--\pageref{LastPage} ~\textbar  \hspace{2pt}\thepage}}}
\fancyfoot[LE]{\footnotesize{\sffamily{\thepage~\textbar\hspace{3.45cm} 1--\pageref{LastPage}}}}
\fancyhead{}
\renewcommand{\headrulewidth}{0pt} 
\renewcommand{\footrulewidth}{0pt}
\setlength{\arrayrulewidth}{1pt}
\setlength{\columnsep}{6.5mm}
\setlength\bibsep{1pt}

\makeatletter 
\newlength{\figrulesep} 
\setlength{\figrulesep}{0.5\textfloatsep} 

\newcommand{\topfigrule}{\vspace*{-1pt}%
\noindent{\color{cream}\rule[-\figrulesep]{\columnwidth}{1.5pt}} }

\newcommand{\botfigrule}{\vspace*{-2pt}%
\noindent{\color{cream}\rule[\figrulesep]{\columnwidth}{1.5pt}} }

\newcommand{\dblfigrule}{\vspace*{-1pt}%
\noindent{\color{cream}\rule[-\figrulesep]{\textwidth}{1.5pt}} }

\makeatother

\twocolumn[
  \begin{@twocolumnfalse}
{\includegraphics[height=30pt]{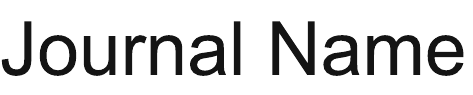}\hfill\raisebox{0pt}[0pt][0pt]{\includegraphics[height=55pt]{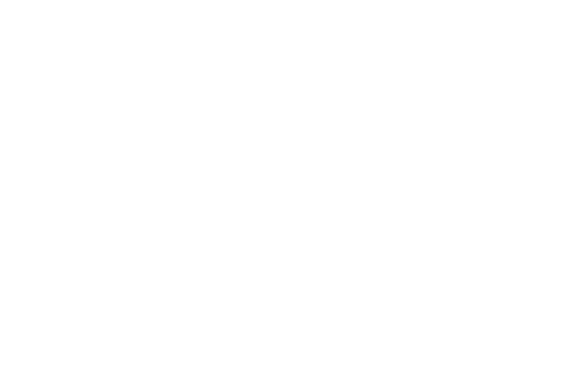}}\\[1ex]
\includegraphics[width=18.5cm]{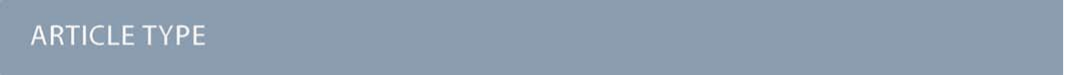}}\par
\vspace{1em}
\sffamily
\begin{tabular}{m{4.5cm} p{13.5cm} }

\includegraphics{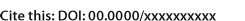} & \noindent\LARGE{\textbf{The ultimate state of elastic turbulence}} \\
\vspace{0.3cm} & \vspace{0.3cm} \\

 & \noindent\large{Piyush Garg\textit{$^{a}$} and Marco Edoardo Rosti\textit{$^{a\dag}$}} \\

\includegraphics{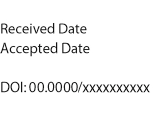} & \noindent\normalsize{The asymptotic properties, with vanishing viscosity, of inertial turbulence in Newtonian fluids have been of intense scrutiny over the years. Much less is known about elastic turbulence - a distinct spatio-temporally chaotic state exhibited by viscoelastic fluids. In this work, using direct numerical simulations of turbulence in a triperiodic box, we show that elastic turbulence achieves an ultimate state with increasing polymer relaxation time, i.e., the Deborah number, for two different constitutive models (Oldroyd-B and FENE-P). In the limiting state, various bulk quantities - the fluid dissipation, the polymeric energy transfer, as well as the elastic stresses - are shown to become independent of the polymer relaxation time scale, notwithstanding the microstructure deformation scales distinctly depending on the model.} \\
\end{tabular}

 \end{@twocolumnfalse} \vspace{0.6cm}

  ]

\renewcommand*\rmdefault{bch}\normalfont\upshape
\rmfamily
\section*{}
\vspace{-1cm}

\footnotetext{\textit{$^{a}$~Complex Fluids and Flows Unit, Okinawa Institute of Science and Technology Graduate University, Okinawa 904-0495, Japan}}

\footnotetext{\dag~Email for correspondence: marco.rosti@oist.jp}

\section{Introduction}
When driven out of equilibrium by a forcing, fluids display complicated turbulent motion dominated by their inertia. Starting with the seminal work of Taylor and Kolmogorov, the limiting state of inertial turbulence with vanishing fluid viscosity (increasing Reynolds number $Re$) has been sought \cite{taylor1935statistical, kolmogorov1991local, batchelor1953theory, frisch1995turbulence, sreenivasan2025turbulence}. Surprisingly, for $Re \rightarrow \infty $ the average fluid dissipation rate remains finite, implying a dissipative anomaly \cite{frisch1995turbulence, sreenivasan2025turbulence, vassilicos2015dissipation, yeung2025small} (although recent results seem to question it \cite{iyer2025whither}). Similar limiting states have been anticipated and searched for in more complex settings involving walls \cite{smits2011high}, convection \cite{lohse2024ultimate}, passive scalars \cite{sreenivasan2019turbulent}, and many more. Most fundamental theories of turbulence in Newtonian fluids are based on describing such states \cite{frisch1995turbulence, sreenivasan2025turbulence, vassilicos2015dissipation}.

Most fluids around us have an internal microstructure which exerts an additional elastic stress in response to external forcing. In the last decades, it has become clear that such viscoelastic fluids exhibit a novel form of turbulence driven by the microstructure, usually termed ``elastic turbulence'' (ET) \cite{steinberg2021elastic}. It was first characterized in detail about two decades ago for flows of polymer solutions with curved streamlines, and was thought to be driven by the elastic hoop stress \cite{steinberg2021elastic, groisman2000elastic, larson2000turbulence, datta2022perspectives, groisman2004elastic, liu2013polymer}. Since its original discovery, however, elastic turbulence has been identified in an increasingly broad range of flow configurations, including homogeneous box turbulence \cite{burghelea2007elastic, berti2008two, berti2010elastic, singh2024intermittency, yerasi2024preserving}, pressure-driven channel and pipe flows \cite{qin2017characterizing, qin2019flow, lellep2024purely, rota2024unified, bonn2011large}, free jets \cite{yamani2021spectral, yamani2023spatiotemporal, soligo2023non}, porous media \cite{haward2021stagnation, browne2021elastic, browne2024harnessing}, and canopy flows \cite{de2023canopy, lopez2025canopy}. More recently, it has also been reported in particle-laden \cite{sun2023anomalous} and bubbly flows \cite{ravisankar2025elastic}, as well as at the small scales of inertial polymeric turbulence \cite{garg2025elastic}. These findings indicate that elastic turbulence is not a peculiarity of a specific flow geometry but rather a generic feature of viscoelastic flows, playing a role analogous to that of inertial turbulence in Newtonian fluids. Intriguingly, over the years, there have been scattered hints in the literature that point to turbulence in viscoelastic fluids becoming independent of the polymeric properties. Prominently, the drag reduction obtained by adding polymers to inertial wall turbulence is known to saturate at the maximum drag reduction asymptote, and does not increase further with polymer concentration \cite{mung08, procaccia2008colloquium, owolabi2017turbulent}. A maximum drag enhancement asymptote has been found also for viscoelastic rotating Couette flows \cite{zhu2023maximum, lin2024maximum}. Very recently it was reported that in experiments of bulk, inertial polymeric turbulence, a state of maximum dissipation reduction is reached with increasing polymer concentration \cite{vxvs-zdbt, rosti2023large}. It would thus appear that increasing elasticity has no effect in these varied limiting states of viscoelastic turbulence; although it should be noted that for most of these cases the inertial effects are not negligible. 

Inspired by the preceding work for inertial turbulence, one is thus tempted to ask the question if there is an ultimate state of elastic turbulence for $De \rightarrow \infty$; the Deborah number $De$ is the non-dimensional measure of the elasticity of the fluid - given by ratio of the microstructure relaxation time scale to the fluid time scale. In addition to clarifying the intrinsic fundamental physics behind ET, such a state would also help in extending existing results from lower $De$ to higher $De$ numbers that are often achieved in experiments but are currently out of the reach of numerical simulations. In this work, we approach this question through direct numerical simulations of turbulence in a triperiodic box at large $De$. Indeed, we find that the bulk properties of elastic turbulence saturate for large $De$ - for two independent constitutive models.  

\section{Methods}
The incompressible velocity $\boldsymbol{u}$ ($\boldsymbol{\nabla} \cdot \boldsymbol{u} =0$) in a viscoelastic fluid is governed by the Navier-Stokes equations,
\begin{equation}
\rho_f \left( \partial_t \boldsymbol{u} +  \left( \boldsymbol{u} \cdot \boldsymbol{\nabla} \right) \boldsymbol{u} \right)=-\boldsymbol{\nabla} p + \mu_f \nabla^2 {\boldsymbol{u}} + \boldsymbol{\nabla} \cdot \boldsymbol{T} + \boldsymbol{F_{ext}}, 
\end{equation}
where $\rho_f$ and $\mu_f$ are the fluid density and dynamic viscosity (with $\nu_f=\mu_f/\rho_f$ the kinematic viscosity), and $F_{ext}$ an external force used to sustain a fully turbulent flow with stationary statistics, made by a combination of sinusoids with a wavelength equal to the domain size and amplitude kept the same in all the directions and for all the investigated cases, that we use to set the Reynolds number of the flow. In particular, the turbulent state is sustained by the Arnold-Beltrami-Childress (ABC) forcing \cite{arnold_2013a} with $A=B=C$. 

In the momentum equation, $\boldsymbol{T}$ is the additional stress due to the polymer molecules, which is dependent on the conformation tensor $\boldsymbol{C}$ describing the state of the microstructure. Here, we consider two distinct constitutive models. The first is the Oldroyd-B model which is well known to reproduce a variety of experimental results \cite{datta2022perspectives}. The model does not limit the extensibility of the polymers, and the stress $\boldsymbol{T}$ is linearly dependent on $\boldsymbol{C}$, i.e., $\boldsymbol{T} = (\mu_p/\tau_p) (f_1 \boldsymbol{C} - f_2 \boldsymbol{I})$ with $f_1 = f_2 = 1$; in the previous relation, $\mu_p$ is the polymeric contribution to the viscosity and $\tau_p$ the relaxation time. The second is the FENE-P model, which sets the maximum extensibility $L_m$ for the polymer molecules by a non-linear relationship between $\boldsymbol{T}$ and $\boldsymbol{C}$: $\boldsymbol{T} = (\mu_p/\tau_p) (f_1 \boldsymbol{C} - f_2 \boldsymbol{I})$ with $f_1 = \frac{L_{m}^2}{L^2_{m}-\mathrm{tr}(\boldsymbol{C})}$ and $f_2 = \frac{L_{m}^2}{L^2_{m}-3}$. In the simulations considered in this work, we fix $L_m = 50$, a value on the shorter end of recent experimental estimates \cite{yamani2021spectral, yamani2023spatiotemporal}. Thus, the choice of these two models ensures that we cover a vast spectrum of extensibility, and other values of $L_m$ can be expected to show an intermediate behavior of the two cases. The governing equation for $\boldsymbol{C}$ is
\begin{equation}
\partial_t \boldsymbol{C} + \boldsymbol{u} \cdot \boldsymbol{\nabla} \boldsymbol{C} = \boldsymbol{C} \cdot \boldsymbol{\nabla}  \boldsymbol{u} + \boldsymbol{\nabla} \boldsymbol{u}^{T} \cdot \boldsymbol{C} - \frac{1}{\tau_p} \left( f_1 \boldsymbol{C}-f_2\textbf{I} \right),
\end{equation}
with $f_1$ and $f_2$ depending on the considered model, Oldroyd-B and FENE-P \cite{bird1987dynamics, larson2013constitutive}. 

The non-dimensional parameters governing the flow are the Deborah number $De = \tau_p/\tau_L$, with $\tau_L$ being the large eddy time scale defined as the ratio of the domain size $L$ (equal to the scale of the external forcing, which is the same for every case considered in this work) and the root mean square (rms) of the velocity field $u_{rms}$ (evaluated for each case), the Reynolds number $Re = \rho u_{rms} L/ \mu_f$, and the viscosity ratio $\beta = \mu_f/(\mu_p+\mu_f)$. In the present work, we set $\beta = 0.9$ for all the results, thus focusing on dilute solutions, and choose the forcing to achieve a $Re \approx 11$ (based on the forcing amplitude and wavelength), smaller than the critical one \cite{podvigina_pouquet_1994a}. While the value of $Re$ is finite, we later show that the effect of inertia is mostly negligible with the present setup. Here, we mostly vary $De$ from $1.25$ to $300$ for the Oldroyd-B cases, and from $1.25$ to $30$ for the FENE-P ones. 

In summary, the simulations reported in this work have been performed using the following parameters in code units: $A=B=C=5$, $L=2\pi$, $\mu_f=0.18$, $\mu_p=0.02$, $\rho=1$, $\tau_p\in \left[ 0.375; 90 \right]$.

\subsection{Numerical discretisation}
The governing equations are solved by the in-house solver \textit{Fujin}\cite{rosti_2026a}, with spatial derivatives discretized using a second order central finite difference scheme, and time integration handled using the Adams-Bashforth scheme, except for the non-Newtonian stress term, which is handled with the Crank-Nicholson scheme. This approach has been applied and validated for viscoelastic flows extensively, see e.g. \cite{rosti2023large, abdelgawad2023scaling, singh2024intermittency}.

$512^3$ grid points are used to discretize the domain for all the cases (with an additional simulation performed with $256^3$ at large $De=100$ to prove that no difference in the results arises), along with the logarithmic formulation of the constitutive tensor, which ensures the positive-definiteness of the conformation tensor by construction at arbitrarily large-$De$ \cite{fattal2004constitutive} and avoids the spurious artifacts introduced by the introduction of artificial stress diffusivity that is often used for numerical stabilization \cite{yerasi2024preserving}. The time-step is kept fixed among all the simulations, and set equal to $2\cdot 10^{-5} u_{rms}/L$.

For all the cases, statistics are obtained by averaging over at least few $t/\tau_L \gg O(De)$ after the statistically stationary state is reached, which implies that the simulations must be carried out for extremely long times at the largest Deborah numbers. In particular, we have run the simulation till $t/\tau_L \sim 1800$, and used the last $600$ time units for averaging the statistics, and we have verified that longer time series do not alter the flow statistics reported in this work. Such averaging time corresponds to twice the Deborah number for the largest $De$ analyzed ($De = 300$), and around 150 times the Deborah number for the smallest $De$ analyzed ($De = 1.25$). Statistical convergence was checked by computing the same statistics with half data points and verifying that they do not changes significantly.

\begin{figure}[t]
	\includegraphics[width=0.49\textwidth]{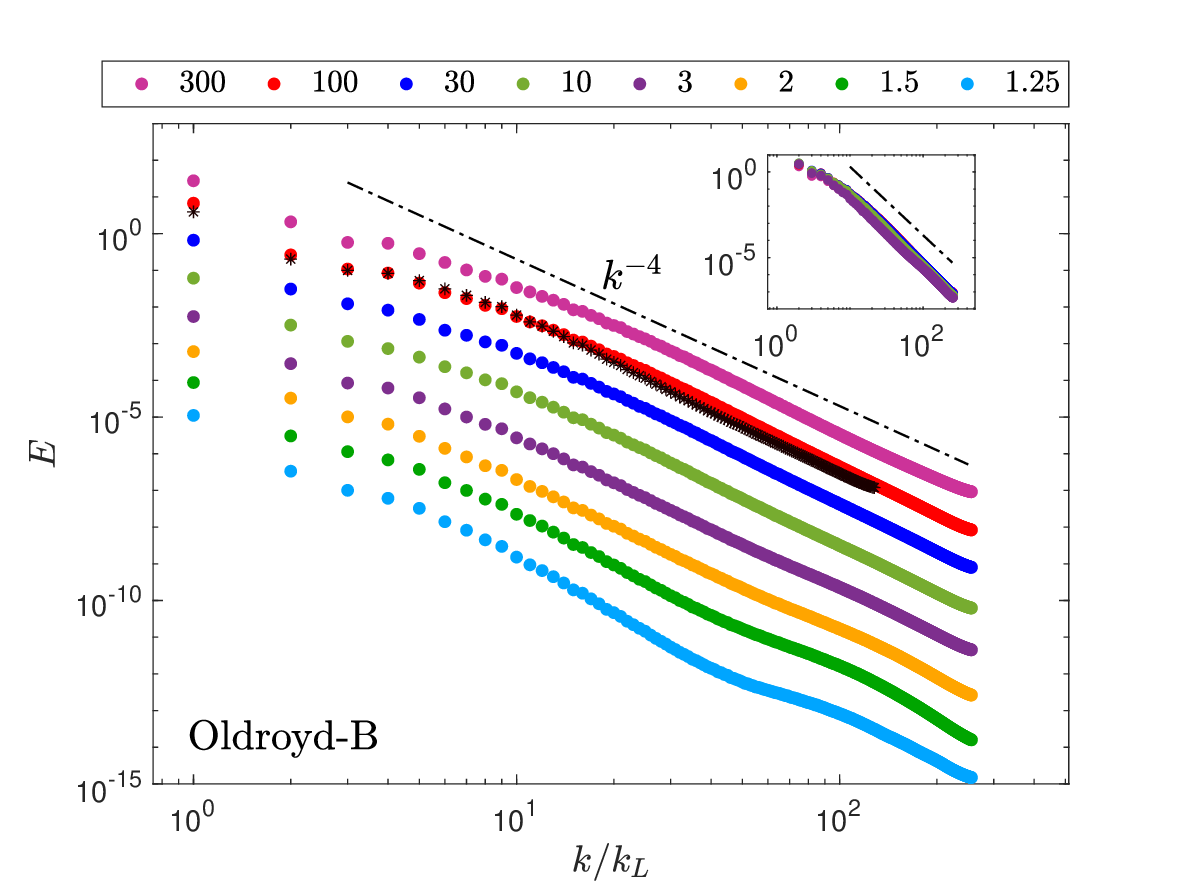}
	\includegraphics[width=0.49\textwidth]{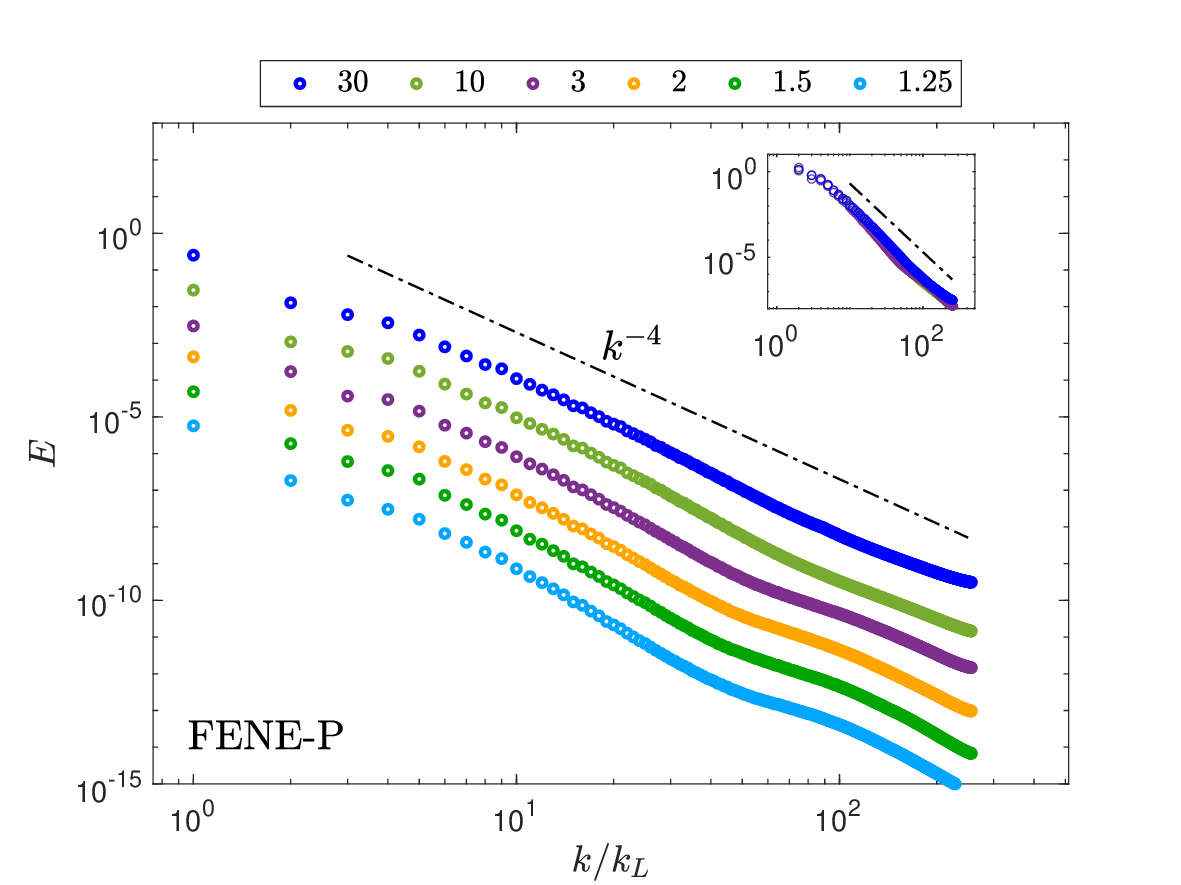}
	\caption{The fluid kinetic energy spectrum $E$ versus the wavenumber $k$, for varying Deborah number $De$, where the cases are vertically shifted for clarity. The insets show the energy spectra without any shift for $De\ge 3$. The curve with black stars in the top panel represents an additional simulation at $De=100$, made with half the grid points in each directions, showing no appreciable differences. The top panel is for the Oldroyd-B model and the bottom one for the FENE-P model.}
	\label{fig1}
\end{figure}

\begin{figure}[t]
	\includegraphics[width=0.49\textwidth]{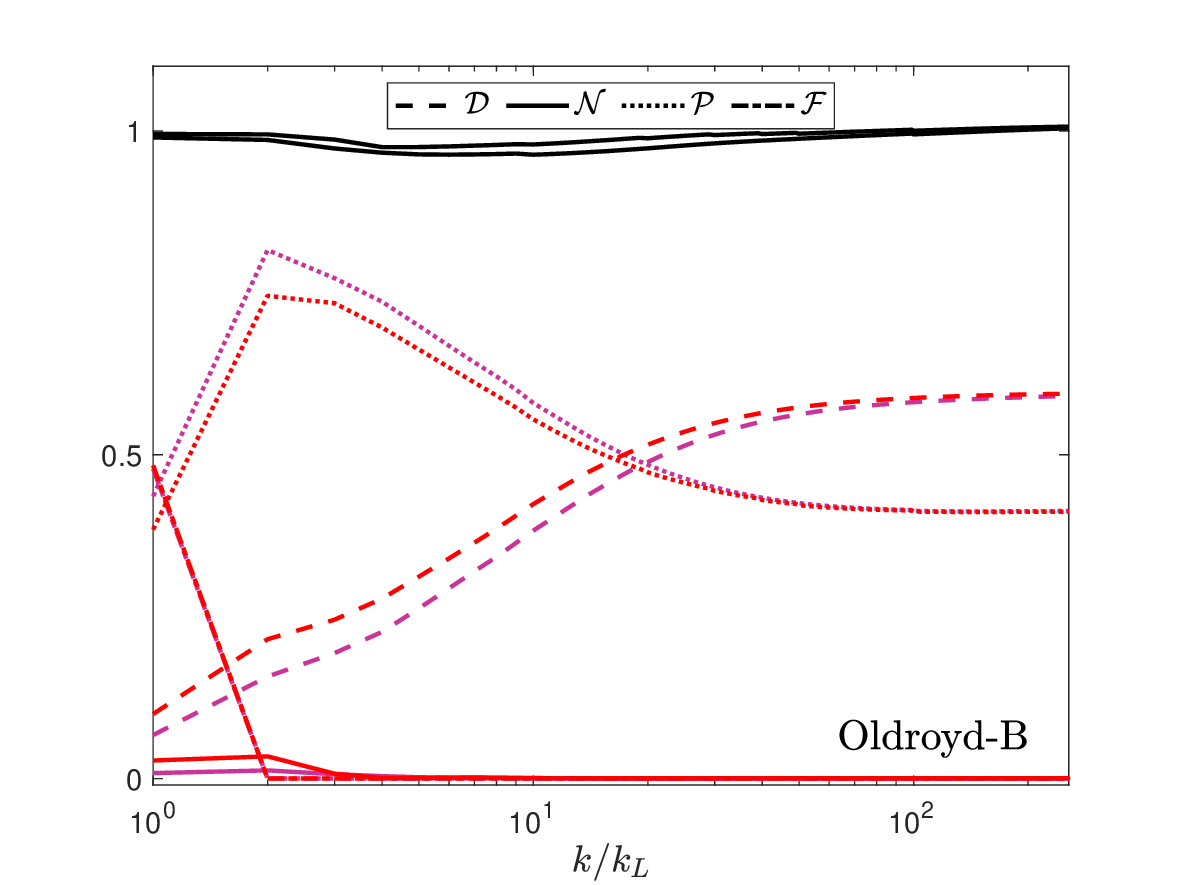}
	\includegraphics[width=0.49\textwidth]{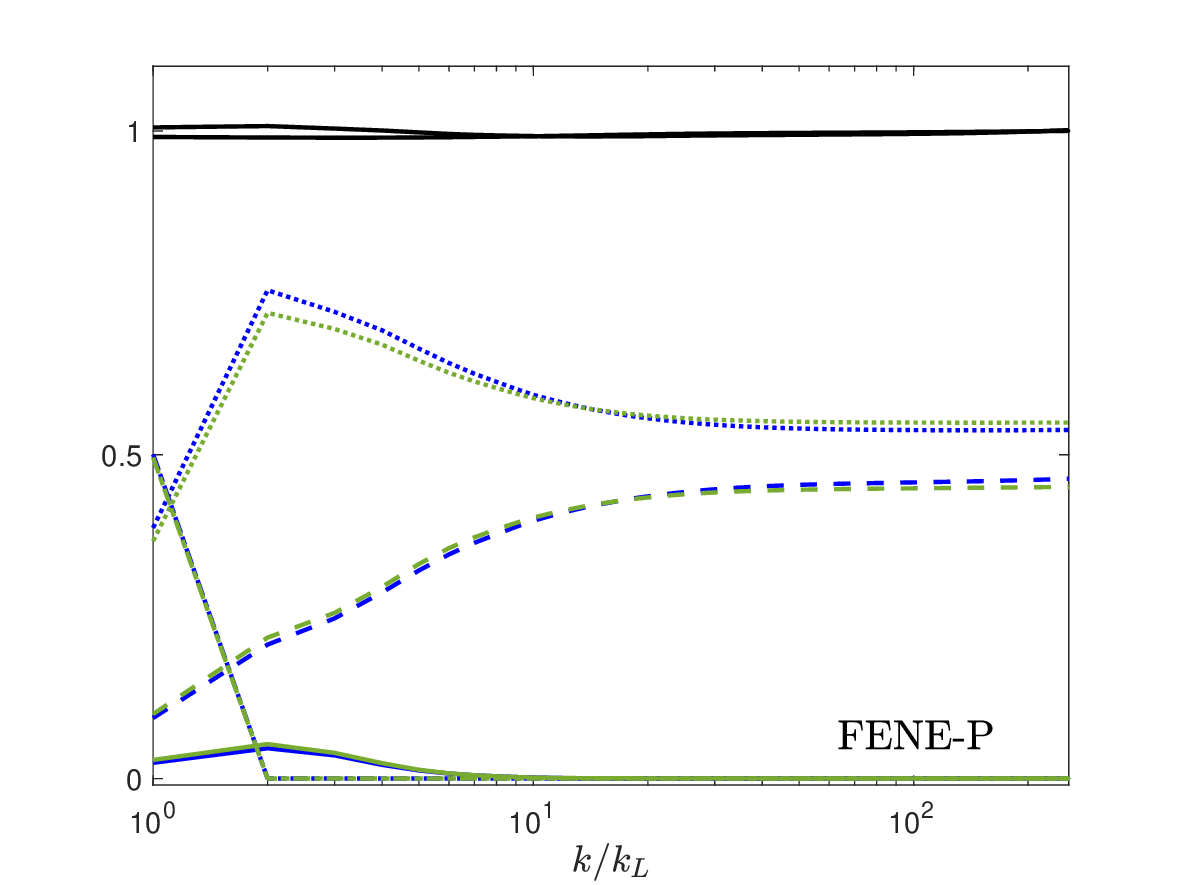}
	\caption{The different contributions to the energy balance of equation~\ref{eq:energy-balance}: viscous dissipation ($\mathcal{D}$), inertial ($\mathcal{N}$),  polymeric ($\mathcal{P}$), and driving force ($\mathcal{F}$) contributions for (top) the Oldroyd-B model ($De=100$ and $300$) and for (bottom) the FENE-P model ($De=10$ and $30$). All terms are normalised by $\varepsilon_f+\varepsilon_p$. The black lines represent the sum of all the contributions, which should sum to one at all wavenumbers. The maximum mismatch found (of the order of $3\%$) is a measure of the accuracy of the numerical method, as well as of statistical convergence.}
	\label{fig2}
\end{figure}

\begin{figure}[t]
	\includegraphics[width=0.49\textwidth]{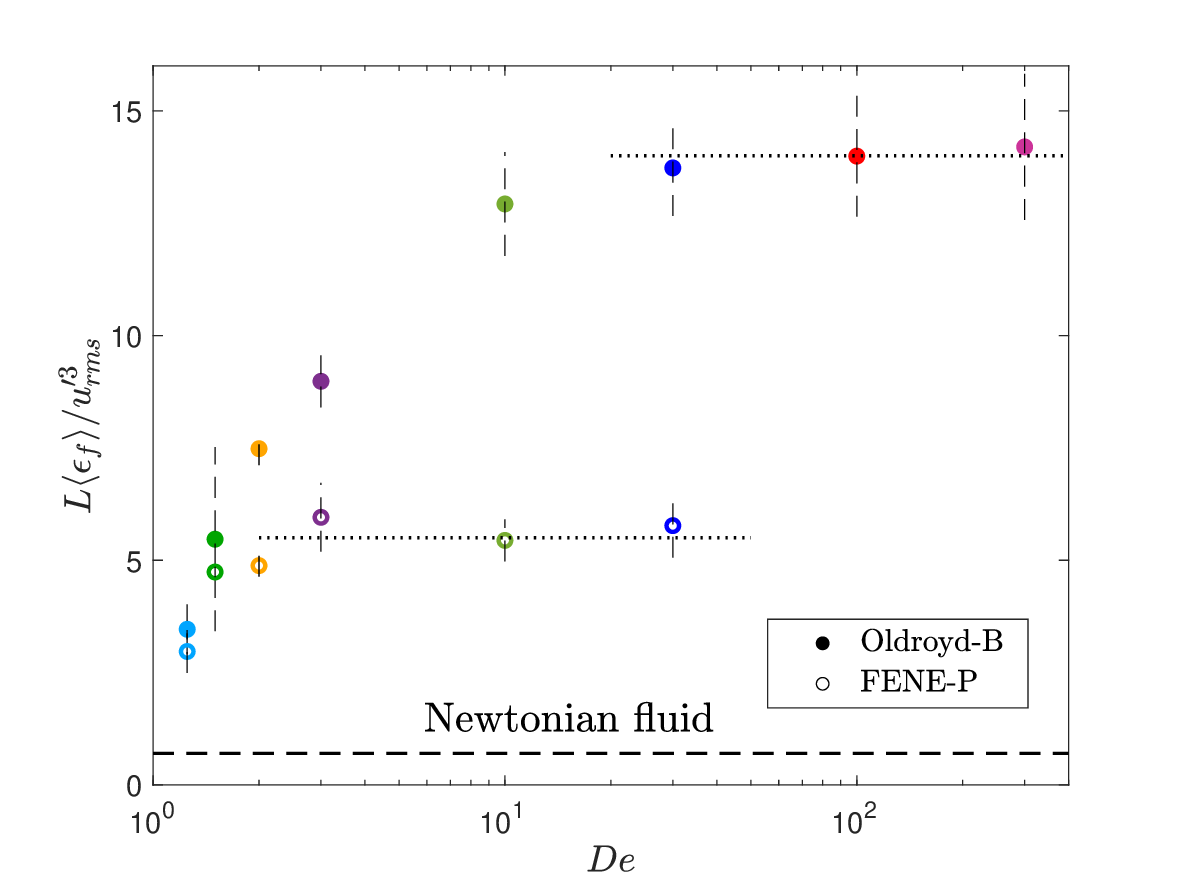}
	\includegraphics[width=0.49\textwidth]{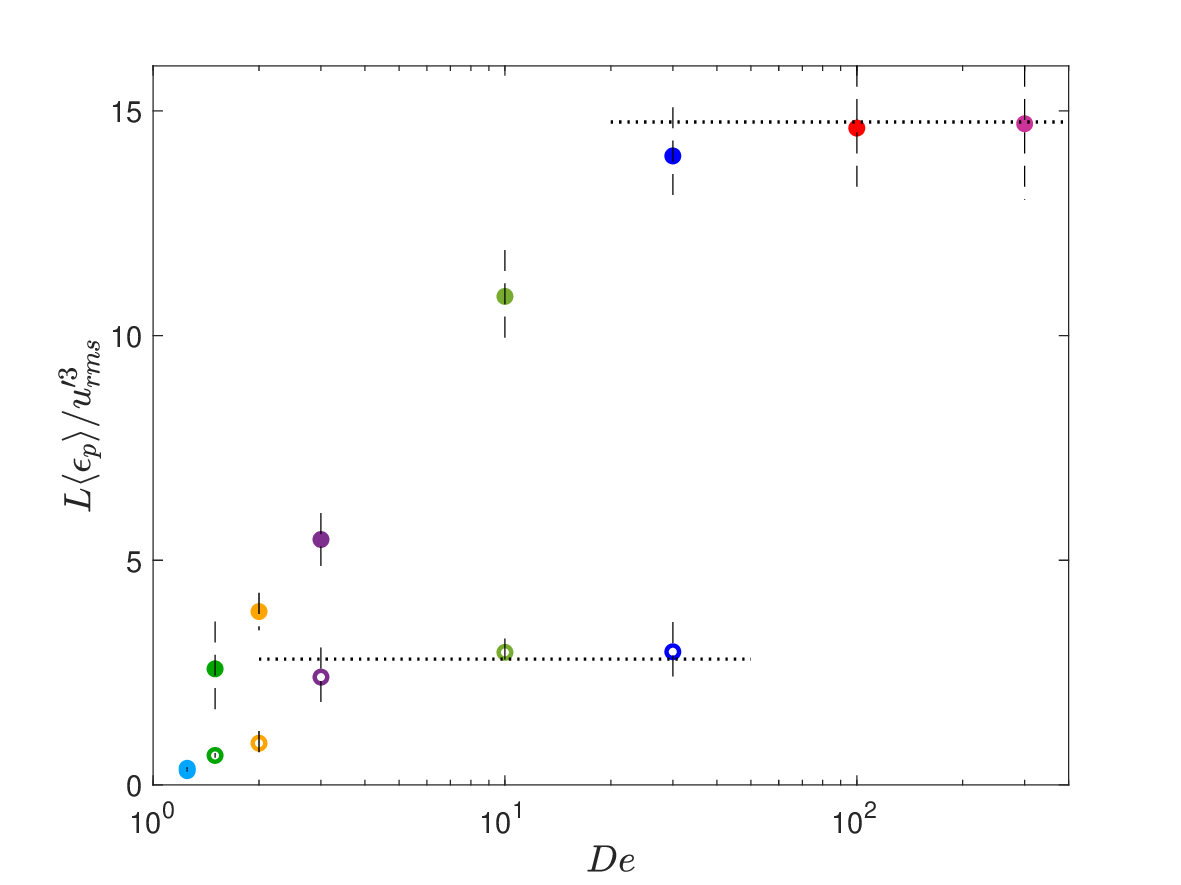}
	\caption{The variation of the non-dimensional (top) fluid dissipation $\epsilon_f$ and (bottom) polymeric energy transfer $\epsilon_p$ with the Deborah number $De$, for the Oldroyd-B and FENE-P models. In the results presented, along with the mean we also show the standard deviation for the various quantities as error bars to indicate the observed variation. The dashed line in the top panel represents the expected value for a Newtonian turbulent flow at the considered $Re$ \cite{doering_foias_2002a}.}
    \label{fig3}
\end{figure}

\section{Results}
Figure~\ref{fig1} shows the fluid kinetic energy spectra versus the non-dimensional wavenumber $k/k_L$ (being $k_L = 2 \pi/L$); the same power-law scaling ($E \sim k^{-4}$) persists till even $De = 300$. The observed scaling of $E(k)$ is consistent with existing experimental observations \cite{steinberg2019scaling, steinberg2021elastic, burghelea2007elastic} and numerical simulations \cite{berti2008two, singh2024intermittency, lellep2024purely, rota2024unified, serafini_2025a}, as well as the limited constraint provided by the only existing theory \cite{fouxon2003spectra}. The energy spectra limit is thus achieved soon after the onset of ET, since a unique scaling range is expected in ET at scales smaller than the forcing, unlike Newtonian turbulence where the extension of the inertial range increases with $Re$. Note however that, the energy spectra of the lowest Deborah numbers exhibit some oscillatory behaviour for $De\le 3$, which can be considered a signature of not having reached yet the fully developed state of ET, as will be shown later in this work. Interestingly, all these observations are the same for both models considered in this work.
 
With the spectra establishing that ET exists till the largest $De$ in our simulations, we turn to multiscale energy balance \cite{pope_2001a} used to explain the main continuations at play. Following the notation in Ref.~\cite{abdelgawad2023scaling}, the balance can be written as follows
\begin{equation} \label{eq:energy-balance}
\epsilon_f+\epsilon_p = \mathcal{D}(k) + \mathcal{N}(k) + \mathcal{P}(k) + \mathcal{F}(k),
\end{equation}
where $\mathcal{D}$ is the viscous dissipation, $\mathcal{N}$, $\mathcal{P}$, and $\mathcal{F}$ are the inertial, polymeric, and driving force contributions, respectively, $\epsilon_f$ is the fluid dissipation rate, and $\epsilon_p$ is the polymeric energy transfer rate. Figure~\ref{fig2} shows the energy balance for the two models at the two largest values of $De$ simulated. As evident from the figure, the nonlinear energy transfer $\mathcal{N}$ is negligible throughout the spectrum, indicating that inertial nonlinear interactions play only a minor role in the spectral energy balance. Nevertheless, $\mathcal{N}$ is consistently larger in the FENE-P simulations than in the Oldroyd-B case, in agreement with previous studies showing that shear-thinning enhances nonlinear inertial effects \cite{amor_soligo_mazzino_rosti_2024a, rosti_2025a}. The forcing term $\mathcal{F}$ is confined to the injection scale and is therefore non-zero only at the forcing wavenumber $k=k_L$. At all remaining scales, the dynamics of elastic turbulence are governed by a near balance between the viscous dissipation $\mathcal{D}$ and the polymer contribution $\mathcal{P}$ consistent with earlier observations \cite{singh2024intermittency}. More importantly, the results reveal that, for both constitutive models, the two largest Deborah numbers collapse onto nearly identical curves, with both $\mathcal{D}/\left( \varepsilon_f + \varepsilon_p \right)$ and $\mathcal{P}/\left( \varepsilon_f + \varepsilon_p \right)$ attaining the same values for different $De$ at $k=k_\mathrm{max}$. Since $\mathcal{D}(k_\mathrm{max})=\varepsilon_f$ and $\mathcal{P}(k_\mathrm{max})=\varepsilon_p$, we therefore wonder whether the fluid and polymer dissipation rates become independent of $De$.

For large $De$, we can expect the turbulent properties to become independent of material properties, if there is a limiting state. To prove this, we use the integral length scale (here approximated by the domain size $L$) and the root mean square velocity of the fluctuations ($u^\prime_{rms}$) to non-dimensionalize all quantities in the results presented, being the most natural choice. Note that, this is also consistent with what done in inertial turbulence, and allows for a direct comparison, especially considering that ET can appear also at large Reynolds numbers when focusing on the smallest scales of the flow \cite{garg2025elastic}. Also, due to the steep decay of the energy spectrum in ET, the integral length scale basically corresponds to the domain size (as well as to the length scale of the external forcing) \cite{taylor_2026a}. Figure~\ref{fig3} shows the variation of the non-dimensional fluid dissipation $\epsilon_f$ and the polymeric energy transfer $\epsilon_p$ with $De$. These are defined as
\begin{equation}
	\epsilon_f  =  \nu_f \langle \nabla  \boldsymbol{u}' \boldsymbol{:} \nabla \boldsymbol{u}' \rangle \nonumber \;\;\;\;\;\; \textrm{and} \;\;\;\;\;\; \epsilon_p  =  \langle \boldsymbol{u}' \cdot \nabla \cdot \boldsymbol{T}' \rangle /\rho_f ,
\end{equation}  
where $\boldsymbol{u}' = \boldsymbol{u} - \bar{\boldsymbol{u}}$ and $\boldsymbol{T}' = \boldsymbol{T} - \bar{\boldsymbol{T}}$ represent the velocity and stress fluctuations, with $\bar{~}$ and $\langle ~ \rangle$ being the time- and space-average operators, respectively. Note that, while the flow has zero time and space averaged mean profiles, the flow has 3D mean profiles when only time averaged. The profiles are the same observed in a laminar flow, except for a modulation in their amplitude (similarly to what observed in the past for the Kolmogorov flow~\cite{musacchio_boffetta_2014a}). Both $\varepsilon_f$ and $\varepsilon_p$ become independent of the Deborah number in the limit of large elasticity ($De\gg1$). Increasing the polymer relaxation time further, and thus strengthening the elastic response, no longer alters the bulk fluid and polymer dissipation sustained by the turbulent fluctuations, pointing to the existence of an \textit{elastic anomaly}. This asymptotic regime is observed for both the Oldroyd-B (figure~\ref{fig3}-top) and FENE-P (figure~\ref{fig3}-bottom) models, indicating that it is robust with respect to polymer extensibility. The asymptotic values of $\varepsilon_f$ and $\varepsilon_p$, however, differ between the two constitutive models, as does the Deborah number at which the asymptotic regime is reached: approximately $De \gtrsim 30$ for Oldroyd-B and $De \gtrsim 3$ for FENE-P. The asymptotic dissipation rates and the corresponding threshold Deborah number are, in general, expected to depend also on the Reynolds number $Re$ and the solvent viscosity ratio $\beta$, although this is not investigated here.

\begin{figure}
	\includegraphics[width=0.49\textwidth]{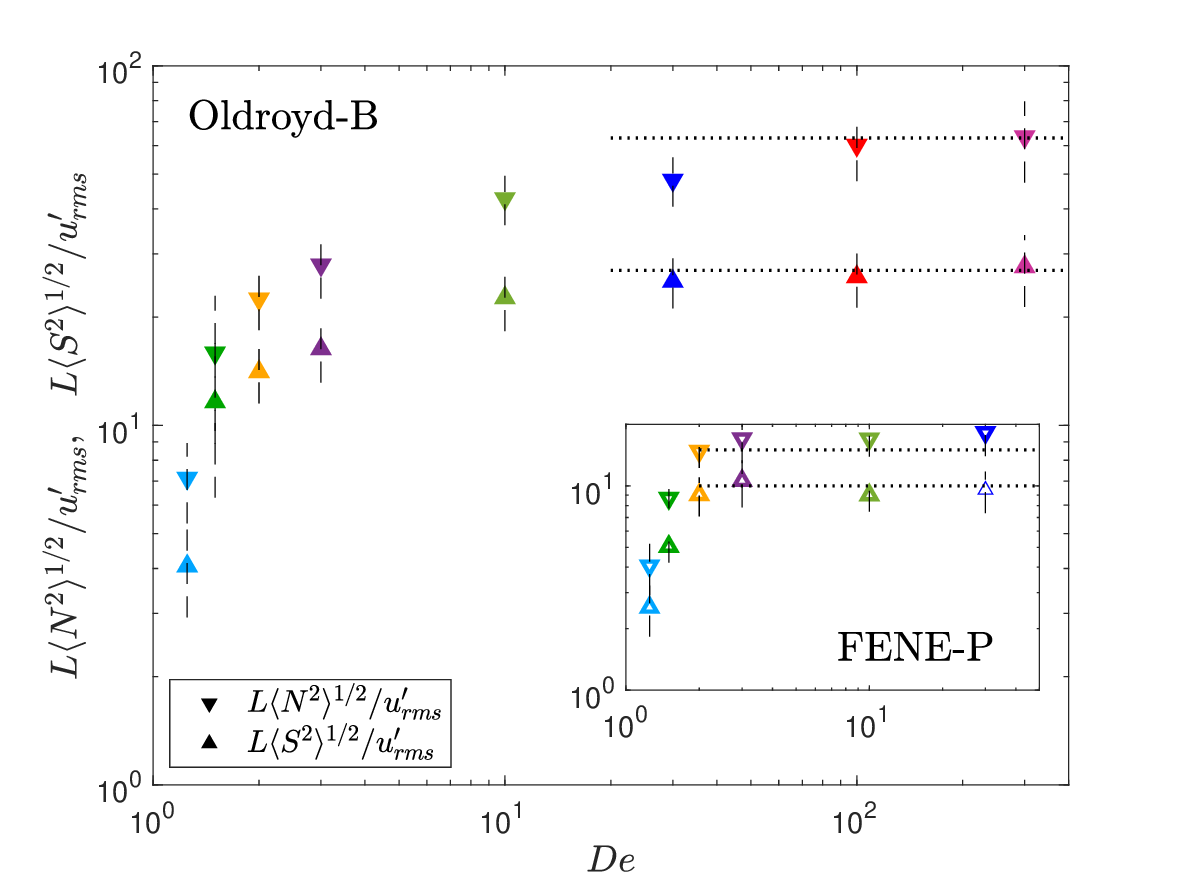}
	\caption{The variation of the non-dimensional root mean square of the polymeric normal stress $N$ and shear stress $S$ with $De$, for the Oldroyd-B model in the main figure and for FENE-P model in the inset.}
    \label{fig4}
\end{figure}

To further prove the existence of the limit, we next turn in figure~\ref{fig4} to the statistics of the polymeric stress tensor $\boldsymbol{T}$, in particular its root mean square. Only two components of the stress tensor $\boldsymbol{T}$ are statistically distinct for isotropic turbulence, and we thus define the normal stress $N=\left( T_{11}+T_{22}+T_{33} \right)/3$ and the shear stress $S=\left(T_{12} + T_{23} + T_{31} \right)/3$. With increasing $De$, initially the elastic stress response increases, but eventually the non-dimensional rms values of both stresses are found to saturate, implying that making the fluid more elastic no longer exerts additional stresses. Even at the largest $De$ we do not find stresses growing in an unbounded manner (even for the Oldroyd-B model), and thus the divergence associated with the coil-stretch transition does not materialize \cite{hinch1977mechanical}, which has been explained before with the feedback of polymers attenuating the stretching action of the velocity gradient, allowing to reach an equilibrium state \cite{balkovsky_fouxon_lebedev_2001a}. The asymptotic regime is attained for both constitutive models, despite their markedly different microstructural dynamics. This is somewhat surprising, as the polymer conformation tensor, $\boldsymbol{C}$, evolves very differently in the two cases.

\begin{figure}[t]
	\includegraphics[width=0.49\textwidth]{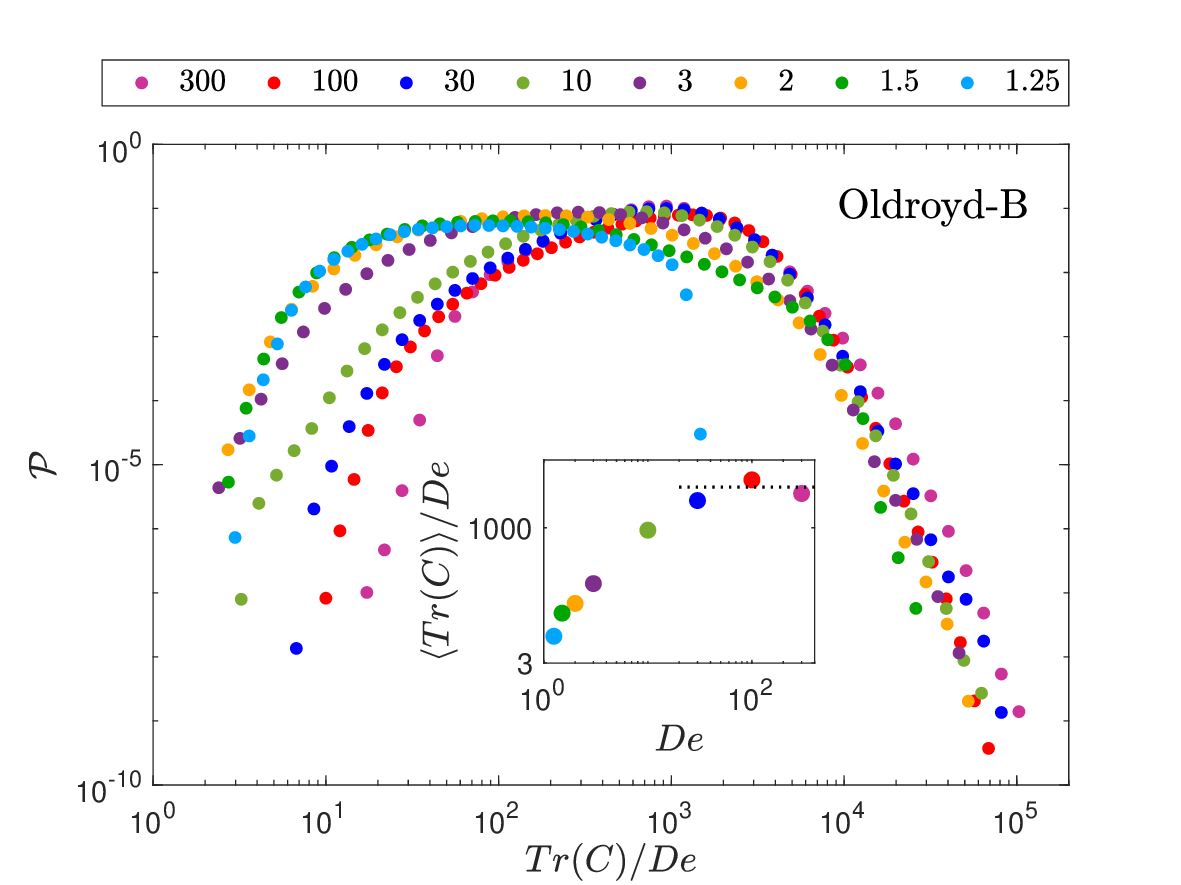}		
	\includegraphics[width=0.49\textwidth]{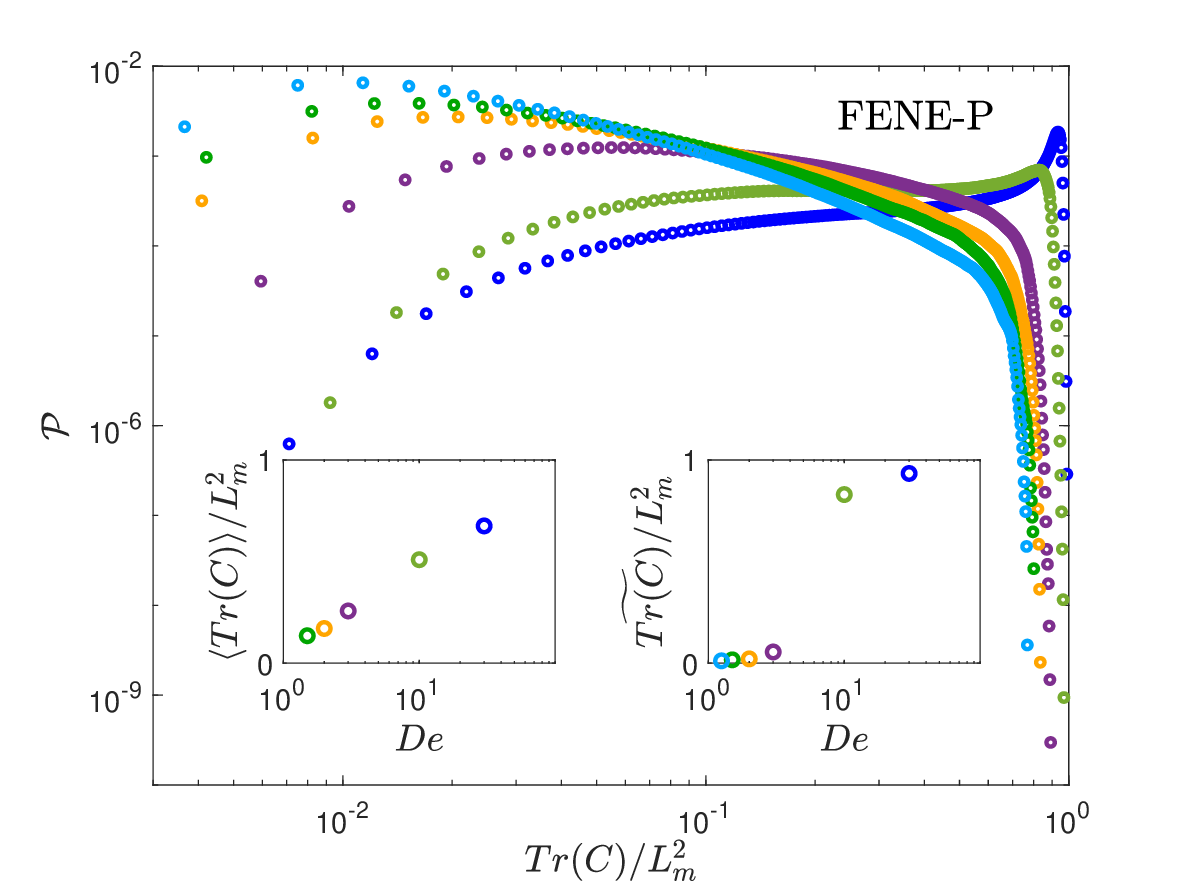}
	\caption{The probability density function of the trace of the conformation tensor (top) $Tr(C)/De$ versus $De$ for the Oldroyd-B model, and (bottom) $Tr(C)/L_m^2$ versus $De$ for the FENE-P model; the insets on the left show the variation of the mean $\langle Tr(C)\rangle$ for both cases, while that on the right of the mode $\widetilde{ Tr(C) }$ for the FENE-P model.}
	\label{fig5}
\end{figure}

\begin{figure}[t]
	\includegraphics[width=0.49\textwidth]{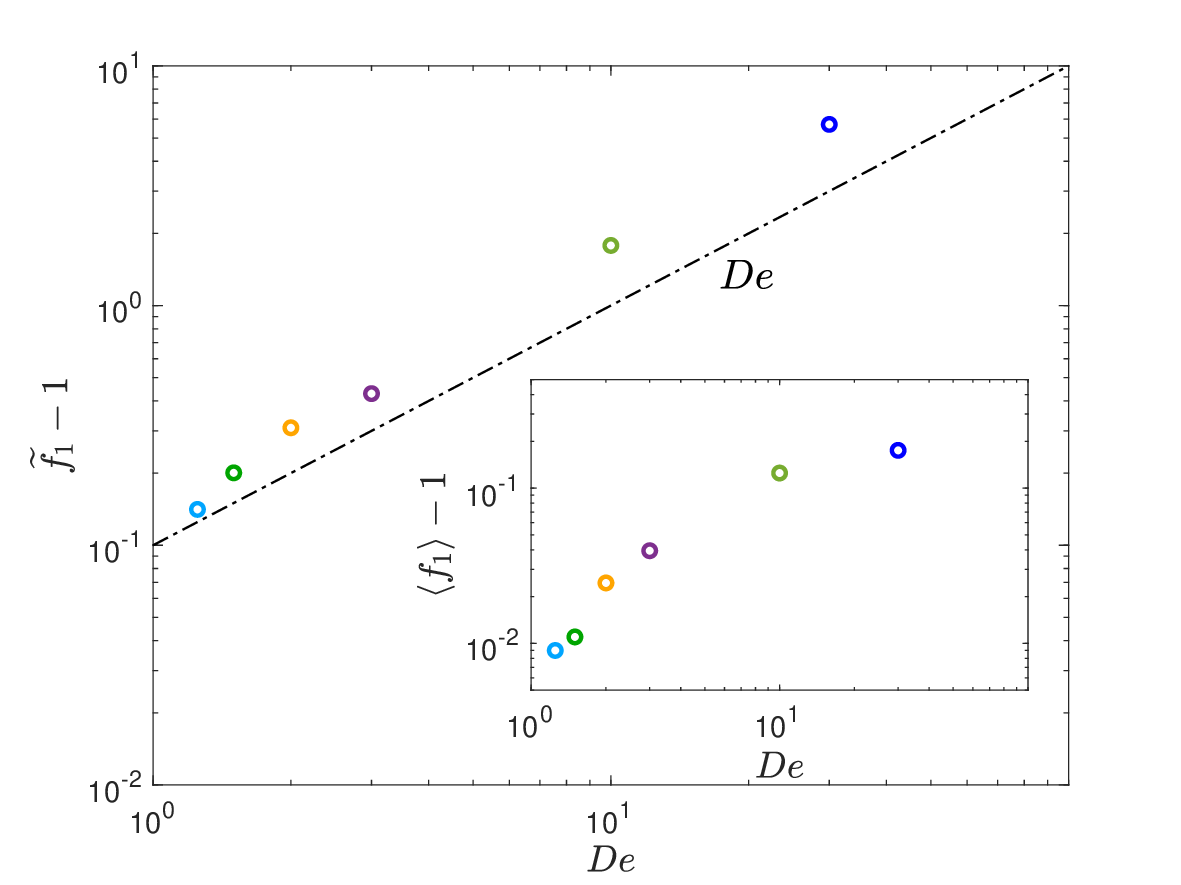}		
	\caption{The mode (main panel) and mean (inset) of the Peterlin fucntion $f_1$ as a function of the Deborah number $De$, for the FENE-P model. The dash-dotted curve is a line with unitary slope.}
	\label{fig6}
\end{figure}

For the Oldroyd-B model, polymer molecules can stretch without bound, and consequently the mean polymer extension grows approximately linearly with the Deborah number, $\langle Tr(C)\rangle\sim De$, as confirmed by the inset of figure~\ref{fig5}-top, where $\langle Tr(C)\rangle/De$ approaches a constant for $De \gtrsim 30$. However, the existence of an asymptotic macroscopic state requires not only the mean extension to scale with $De$, but also the fluctuations around the mean to exhibit the same scaling. Indeed, the right tail of the probability density function of $Tr(C)$, corresponding to the most strongly stretched polymers, collapses when rescaled by $De$ (figure~\ref{fig5}-top). Since the polymer stress $\boldsymbol{T}$ in the Oldroyd-B model is proportional to $\boldsymbol{C}/De$, this scaling directly implies that the polymer stress reaches a limiting distribution, consistent with the saturation of the dissipation rates observed above.

The mechanism is fundamentally different for the FENE-P model. Here, polymer extension is bounded by the finite extensibility $L_m$, which truncates the probability distribution of $Tr(C)$ and strongly modifies its tail (figure~\ref{fig5}-bottom). For $De \gtrsim 3$, the mode of the distribution reaches $L_m^2$, indicating that most polymer molecules have already attained their maximum extension beyond the coil--stretch transition. Similar shifts toward maximal extension have previously been reported for polymers in inertial turbulence \cite{watanabe2010coil, picardo2023polymers} and random flows \cite{liu2010stretching, musacchio2011deformation}; here we show that the same phenomenon also occurs in elastic turbulence. Past works of polymer dynamics in random flows \cite{balkovsky2001turbulence, gerashchenko_chevallard_steinberg_2005a, watanabe2010coil}   suggested that $\langle Tr(C)\rangle$ saturates to values smaller than $L_m^2$ as $De \rightarrow \infty$; our data in the left inset of figure~\ref{fig5}-bottom do not show any sign of saturation within the observed range. However, as $De \rightarrow \infty$, the mode of the distribution $\widetilde{Tr(C)}$ approaches $L_m^2$ (i.e., $\widetilde{Tr(C)} \rightarrow L_m^2$ as shown in the right inset of figure~\ref{fig5}-bottom), and thus the conformation tensor can remain bounded as an increasing fraction of polymers become fully stretched. Despite this distinct dynamics of $\boldsymbol{C}$, also the polymer stress $\boldsymbol{T}$ can attain a limiting state in the FENE-P model. This is made possible by the Peterlin function $f_1$, which modifies the constitutive relation between $\boldsymbol{C}$ and the stress tensor $\boldsymbol{T}$. As $\widetilde{Tr(C)}$ approaches $L_m^2$, the Peterlin function increases approximately linearly with $De$ (figure~\ref{fig6}), compensating for the bounded polymer extension and allowing the stress to saturate when $De \rightarrow \infty$. Importantly, this asymptotic regime is reached well before every polymer molecule is fully extended.

Note that, the small $L_m$ was deliberately chosen to understand dynamics not captured by the Oldroyd-B model; the behavior of polymers with a finite but large $L_m$ can be expected to be a mixture of the two limiting behaviors discussed here, with the ultimate state of ET arising regardless of the value of $L_m$.

\section*{Conclusions}
In this work, we have shown that for large $De$, elastic turbulence reaches an ultimate state where the bulk properties, most importantly fluid dissipation and polymeric energy transfer, become independent of the material properties of the polymers. Once the ultimate state is reached, increasing the elasticity no longer affects the induced turbulence. While our analysis is limited to a single value of Reynolds number $Re$ and polymer concentration $\beta$, and also by a rather limited range of saturated $De$ (the maximum currently achievable), crucially, we have shown this limiting state for two distinct constitutive models (the Oldroyd-B and FENE-P models), and hence it should be expected to exist in a range of viscoelastic fluids which exhibit ET, although with different asymptotic values. In both the considered models, the ultimate state of ET corresponds to the achievement of an added stress $\boldsymbol{T}$ independent of the Deborah number $De$, and as a consequence all fluid properties are independent of $De$ too in this ultimate regime. However, such state has significantly different inner polymer dynamics (exemplified by the conformation tensor $\boldsymbol{C}$) depending on the model. In the Oldroyd-B model, the ultimate state originates from the conformation tensor diverging with $De$ being balanced by the diverging $1/\tau_p$ in the stress definition. In the FENE-P model, the conformation tensor is limited by the maximum extensibility $L_m$, so the diverging $1/\tau_p$ in the stress definition is balanced with the diverging Peterlin function $f_1$.

In practical applications of ET, the $De$ number achieved can often exceed that achievable in numerical simulations, and hence the limiting state can be used for predicting quantities of practical interest, such as the mixing efficiency or excess dissipation \cite{Scholz_2014, browne2024harnessing, browne2021elastic}. The limiting state should also be more readily amenable to a theoretical analysis, which remains limited to a few early works for ET \cite{fouxon2003spectra, balkovsky2001turbulence}. 

Existing experimental work on limiting states of turbulence in viscoelastic fluids has mostly been confined to large $Re$, pressure-driven inertial turbulence of viscoelastic fluids in the context of drag reduction \cite{mung08}. A recent experimental study of bulk, polymeric turbulence found the maximum dissipation reduction regime at large $De$ \cite{vxvs-zdbt}. As we recently showed, the small scales of large-$Re$ polymeric turbulence indeed correspond to ET \cite{garg2025elastic}, and thus, it is tempting to suggest that these experimental observations are a manifestation of the ultimate state of elastic turbulence. However, the large-$De$ behavior at large $Re$ is known to be non-trivial and distinct at the large and small scales \cite{rosti2023large, singh2024interplay, garg2025elastic} and further work is needed to firmly establish how the ultimate state is modified with increasing $Re$.

Finally, these results establish that limiting states of turbulence need not be confined to the large inertia limit as often presumed, exemplified by the fact that prior work on even viscoelastic fluids has focused on the large-$Re$ limit. In fact, it is likely that asymptotic states are observable in a whole range of spatiotemporally chaotic states in fluids. Beyond viscoelasticity, the past two decades have seen the proliferation of work in characterizing turbulent states in biological fluids \cite{marchetti2013hydrodynamics} and swimmer suspensions \cite{koch2011collective} - where ``active turbulence'' is driven by energy injection at the micro-scale in contrast to the usual studies of turbulence. Our work opens up the possibility that there is an universal limiting state for active turbulence itself where the properties no longer depend on the detailed activity \cite{marchetti2013hydrodynamics, koch2011collective, alert2022active}, as well as other forms of particle and bubble-laden turbulence \cite{mathai2020bubbly, brandt2022particle, marchioli_rosti_verhille_2025a}. 


\section*{Conflicts of interest}
There are no conflicts to declare.

\section*{Data availability}
The code used for the present research is a standard direct numerical simulation solver for the Navier–Stokes equations. Full details of the code used for the numerical simulations are provided in the text and in Ref.~\cite{rosti_2026a} references. All data needed to evaluate the conclusions are presented in the paper; data of the figures are openly available at \url{https://www.oist.jp/research/research-units/cffu/publications/publication-data}.

\section*{Acknowledgements}
The research was supported by the Okinawa Institute of Science and Technology Graduate University (OIST) with subsidy funding to M.E.R. from the Cabinet Office, Government of Japan. M.E.R. also acknowledges funding from the Japan Society for the Promotion of Science (JSPS), grant 24K17210 and 24K00810. The authors acknowledge the computer time provided by the Scientific Computing \& Data Analysis section of the Core Facilities at OIST, and by HPCI, under the Research Project grants hp250035, hp260009, hp260019, and hp260317. P.G. passed away on 24 October 2025 before submitting this manuscript; this work is dedicated to his memory.



\balance


\bibliographystyle{rsc} 
\providecommand{\noopsort}[1]{}\providecommand{\singleletter}[1]{#1}%
\providecommand*{\mcitethebibliography}{\thebibliography}
\csname @ifundefined\endcsname{endmcitethebibliography}
{\let\endmcitethebibliography\endthebibliography}{}

\end{document}